\documentclass{aastex}          
\usepackage{spr-astr-addons}    

\begin{document}
%
\title{On the Progenitors of Core-Collapse Supernovae}

\shorttitle{Core-Collapse Supernova Progenitors>}
\shortauthors{Leonard}

\author{Douglas C. Leonard\altaffilmark{1}} 

\altaffiltext{1}{Department of Astronomy, San Diego State University, San Diego, CA
  92182, USA}

\begin{abstract}
Theory holds that a star born with an initial mass between about 8 and 140
times the mass of the Sun will end its life through the catastrophic
gravitational collapse of its iron core to a neutron star or black hole.  This
core collapse process is thought to usually be accompanied by the ejection of
the star's envelope as a supernova.  This established theory is now being
tested observationally, with over three dozen core-collapse supernovae having
had the properties of their progenitor stars directly measured through the
examination of high-resolution images taken prior to the explosion.  Here I
review what has been learned from these studies and briefly examine the 
potential impact on stellar evolution theory, the existence of ``failed
supernovae'', and our understanding of the core-collapse explosion mechanism.

\end{abstract}

\keywords{stellar evolution, core-collapse supernovae}

\section{Introduction} 

Which stars explode as core-collapse supernovae (CC SNe)?  Standard theory
suggests that isolated stars with initial masses $\lesssim 8\ M_\odot$ end
non-explosively by forming white dwarfs. Those born with $\gtrsim 8\ M_\odot$
die by exploding.  The vast majority of these massive stars --- those born with
masses between roughly $8\ M_\odot$ and $140\ M_\odot$ --- are believed to die
as CC~SNe through the implosion of their iron cores and the subsequent ejection
of their envelopes, leaving behind neutron stars or black holes.  Although not
yet observationally demonstrated, it is possible that {\it some} of these
massive stars fail to turn implosion into explosion and collapse to a compact
object with no associated ``fireworks'' \citep[i.e., ``failed
  SNe'';][]{Kochanek08}.  Beyond $\sim 140\ M_\odot$ (provided Nature actually
mints, or minted, such stars; e.g., \citealt{Figer05}) death likely arrives
earlier in life through the ``pair-instability'' process that triggers
explosive fusion of the oxygen core and the complete disruption of the star,
resulting in a ``pair-instability supernova''
(\citealt{Rakavy67,Galyam09a,Moriya10}).

Here we focus specifically on CC SNe, what we have learned about the properties
of their progenitor stars through observation, and the implications of these
findings on stellar evolution theory, the existence of failed SNe, and the CC
explosion mechanism.  By registering pre-SN and post-SN images, usually taken
at high resolution using either space-based optical detectors --- or more
recently with ground-based infrared detectors equipped with laser guide star
adaptive optics systems (LGS-AO) --- over three dozen CC SNe have now had the
properties of their progenitor stars either directly measured or constrained by
establishing upper limits on their luminosities.  These studies have enabled
direct comparison with stellar evolution models that, in turn, permit estimates
of the progenitor stars' physical characteristics to be made.  As we shall see,
initial results of these progenitor studies have matched theoretical
expectations in some regards, but the field is young and strewn with hints that
in some areas a rethinking of standard stellar evolution theory may be in
order.

This paper is organized as follows.  \S~2 provides a brief review of CC SN
classification and stellar evolution theory and sketches out generic
expectations for the progenitors of each of the major CC SN types.  \S~3
confronts expectations with existing observations, and \S~4 concludes with a
brief summary and discussion.

\section{Expectations}
\subsection{CC SN Classification, Stellar Evolution, and Zeroth-Order
  Progenitor Expectations}

It is typical to subdivide CC~SNe into at least five major categories (see
\citealt{Filippenko97} for a thorough review): II-Plateau (II-P; hydrogen in
spectrum and plateau in optical light curve), II-Linear (II-L; hydrogen in
spectrum, no plateau in optical light curve), IIn (hydrogen in spectrum and
spectral and photometric evidence for interaction between SN ejecta and a dense
circumstellar medium [CSM]), IIb (hydrogen in spectrum initially, with
transformation into a hydrogen-deficient spectrum at later times), and Ib/c (no
evidence for hydrogen in spectrum at any time).  This ordering is thought to be
a roughly increasing one in terms of inferred degree of envelope stripping
prior to explosion.  That is, SNe~II-P are the least stripped at the time of
explosion and SNe~Ib/c are the most stripped, with the others falling in
between.  Although considerable uncertainty persists regarding the exact
proportions, SNe II-P probably account for $\sim 60\%$ of all CC SNe, SNe~Ib/c
for $\sim 30\%$, and the final $\sim 10\%$ comprised of the rarer IIn, II-L,
and IIb types \citep[e.g., ][and references therein]{Smartt09a}.

The most basic expectation for the stellar progenitors of CC SNe is that they
should have properties consistent with the evolutionary endpoints of stars born
with more than $\sim 8\ M_\odot$, the theoretical lower limit for which
core-collapse will occur.  The expected characteristics of such stars are most
easily viewed on the theoretical Hertzsprung-Russell Diagram (HRD), an example
of which is shown in Figure~1.  Quick examination yields simple predictions:
Stars between about $8\ M_\odot$ and $25\ M_\odot$ should end their lives with
properties consistent with red supergiants (RSG), while those above $\sim
25\ M_\odot$ should end as hot Wolf-Rayet (WR) stars, which have lost all, or
nearly all, of their H and (sometimes) He envelopes.

\begin{figure}[!ht]
\centering
\rotatebox{-90}{
\includegraphics[width=2.5in]{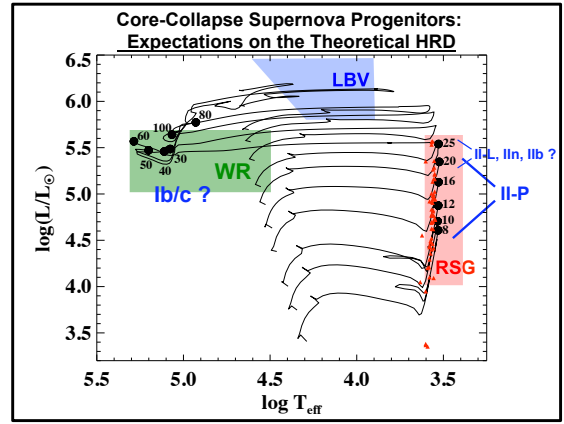}}
\caption{Theoretical HRD with evolutionary tracks ({\it thin lines}) and
  stellar endpoints ({\it large, filled circles}) for solar metallicity stars,
  taken from the STARS stellar evolution models \citep{Eldridge04}.  The models
  follow stellar evolution up to the initiation of core neon burning, which is
  likely to give an accurate indication of the pre-SN luminosity; note that the
  $8\ M_\odot$ model does not include the uncertain ``second dredge-up'' phase,
  which would make its evolutionary endpoint significantly redder and more
  luminous \citep{Eldridge04}.  The {\it RSG} location is indicated along with
  the effective temperatures and luminosities derived for a selection of Milky
  Way RSGs ({\it small, filled triangles}) by \citet{Levesque05}.  The {\it
    LBV} and {\it WR} regions (from \citealt{Smith04a} and \citealt{Smartt09a},
  respectively) are also shown.  Regions from which one might expect SNe~II-P,
  II-L, IIn, IIb, and Ibc to arise from the simple considerations discussed in
  the text (\S~2.2) are also indicated.}
\end{figure}

Stars born with $\gtrsim 50\ M_\odot$ are believed to experience a ``luminous
blue variable'' (``LBV'') phase {\it en route} to their deaths as
stripped-envelope WR stars, during which violent episodic bursts of mass-loss
can occur (eta-carinae being the most famous nearby example).  With
luminosities $\gtrsim 10^6\ L_\odot$, LBVs are the most luminous single stars
known.  An important point is that it has {\it not} traditionally been believed
that such stars will explode as SNe during the LBV stage, since this
evolutionary phase is thought to occur while the star is still at the end of
core H burning, or possibly the beginning of core He burning \citep{Maeder94}.

The bottom line from conventional theory therefore is that CC SNe should arise
from the RSG and WR regions on the HRD.  Naturally, many issues complicate this
simple picture.  First and foremost, the HRD shown in Figure~1 is for single
stars.  Mass transfer with a companion can drastically affect a star's final
properties prior to core collapse, and some fraction of CC~SN progenitors are
surely arising from interacting binary systems \citep[e.g.,][]{Filippenko91,
  Nomoto95}.  Second, although for stars that become RSG it is widely held that
the greater the initial mass the stronger the mass-loss should be, details are
not well established.  There may well be a region in the upper-mass regime of
RSG where stars have lost a significant fraction (but not all) of their H
envelopes prior to exploding.  Some models suggest that such stars could
undergo ``blue loops'' on the HRD, during which they experience temporary
blueward excursions (e.g., \citealt{Xu04}), and it is possible that some of
these stars could even explode during these brief migrations.  This would give
their progenitors characteristics consistent with hotter supergiant stars,
e.g., yellow supergiants (YSG) or, even, blue supergiants (BSG).  In fact, the
BSG progenitor of SN~1987A is sometimes explained in this manner, although
binary scenarios remain very popular for it, as well (e.g.,
\citealt{Morris07}).  Finally, metallicity and treatment of stellar rotation
affect the final stellar characteristics, most notably the dividing line
separating stars that will end as RSGs and those that will end as WRs.  Values
for this cutoff range from $\sim 22\ M_\odot$ to $\sim 34\ M_\odot$, depending
on the model's characteristics (e.g., \citealt{Heger00, Meynet00, Heger03}).
Such caveats aside, the theoretical picture remains fairly clear on the basic
fact that exploding, single stars can be expected to largely populate the RSG
and WR regions of the HRD.

\subsection{Predictions for Specific CC SN Types}

Of all of the types of CC SNe, perhaps the most confidence can be placed on the
predicted SN~II-P progenitor.  SNe~II-P are characterized by an enduring period
($\sim 100$ days) of nearly constant optical luminosity (i.e., a ``plateau'')
and strong spectral evidence for hydrogen at all times.  These properties have
long been thought to result from having the shock-deposited and radioactive
decay energy of the explosion injected into a massive and extended envelope,
which then slowly releases the energy as the hydrogen recombines during the
``photospheric'' phase of the SN's evolution \citep{Chevalier76}.  SNe~II-P are
also very weak radio sources, indicating little interaction with circumstellar
material.  These outstanding characteristics finger single RSG as the likely
progenitors of SNe~II-P.

Clear-cut predictions for the other CC SN types are more difficult to make, but
some reasonable expectations can be proposed based on their observed
characteristics.  Of primary significance is that none exhibit a light curve
that is ``held up'' like an SN~II-P light curve is.  This implies that these
other CC~SNe do not have similarly extended envelopes when they explode.  They
are also typically much stronger radio emitters at early times (suggesting
interaction with CSM) and generally show less evidence for hydrogen in their
spectra (or, in the case of SNe Ib/c, no evidence for hydrogen).  These
characteristics indicate more substantial pre-SN mass-loss (or,
mass-stripping). Putting these clues together, the regions from which we might
expect these events to arise are {\it perhaps} the upper mass regions of the
RSG for SNe~II-L/IIb/IIn, and WR for SNe~Ib/c.

Before proceeding to the observations, we note that a simple calculation finds
that if one simultaneously posits a standard Salpeter IMF (slope $\alpha =
-2.35$), SNe II-P arising from $8\ M_\odot$ to $20\ M_\odot$ progenitors,
SNe~II-L/IIn/IIb resulting from $20\ M_\odot \rightarrow\ 25\ M_\odot$
progenitors, and SNe~Ib/c coming from $25\ M_\odot \rightarrow\ 140\ M_\odot$
progenitors, relative frequencies of $72\%$ (SNe~II-P), $8\%$ (SNe~II-L, IIn,
and IIb), and $20\%$ (SNe~Ib/c) result.  Given the large uncertainties involved
in any attempt to connect such values with observed SN rates (e.g., the strong
sensitivity to the mass cutoffs and the lower mass for which CC is assumed to
occur; the substantial uncertainty in the calculation of relative SN rates; the
effects of binarity; the uncertainty in the IMF slope itself), it is somewhat
reassuring that such a calculation does not produce values {\it wildly}
discrepant with estimates of CC~SN fractions.  While such indirect statistical
comparisons are potentially illuminating, they are no substitute for results
obtained from the direct observations of stars that actually explode as CC~SNe,
the subject to which we now turn.

\section{Results}
\subsection{Experimental Method}

To avoid source confusion, searches for SN progenitors demand two primary
elements: (1) A very nearby SN (generally, within $\sim 20$ Mpc), and (2) a
high resolution pre-SN image (usually taken with the {\it Hubble Space
  Telescope} [{\it HST}]).  If both conditions are met, an additional high
resolution image must be obtained with the SN still visible (but, not saturated
in the image), to permit registration between the pre- and post-SN images to
better than $\sim 30$ milli-arcseconds.  This second image can be obtained at
either optical wavelengths using {\it HST} several months after explosion (so
that the SN is faint enough to permit a deep image) or at near-infrared (NIR)
wavelengths using LGS-AO, which can be obtained nearly immediately (since CC
SNe are quite faint in the NIR, even at early times).  If nothing is found at the SN location in the pre-SN
image, an upper-luminosity limit for the progenitor may be derived from the
image's detection limit.

If spatial coincidence exists between the SN and a point source in the pre-SN
image, the most convincing way to demonstrate its connection to the SN is to
confirm its absence in an image taken years later, after the SN has faded
beyond detection.\footnote{Dust obscuration remains a difficult possibility to
  definitively exclude -- e.g., if substantial dust is formed in the SN
  atmosphere and the putative progenitor star lies behind the SN along the
  line-of-sight, the star could be obscured in post-SN images.}  To date,
roughly one dozen CC SN progenitors have now been directly detected (i.e.,
shown to be spatially coincident with the SN) in pre-SN images, two dozen upper
limits derived from non-detections, and four progenitors confirmed through
their absence in images taken after the SN has faded.  For an exhaustive
listing and discussion of all studies completed through early 2009, see
\citealt{Smartt09a}; details of the techniques used to register images and
derive the upper limits for non-detections may be found in \citet{Leonard18}
and \citet{Leonard20}.

\subsection{Type II-Plateau}

Given their relative frequency, it is not surprising that SNe~II-P are by far
the most well-defined category of CC~SNe in terms of direct observational
progenitor detections and constraints, making up over half of all progenitor
studies to date.  At present, seven putative SN~II-P progenitor detections have
been made using pre-SN images, 12 upper luminosity limits have been derived
from non-detections, and one progenitor has been definitively identified
through its disappearance in post-SN images.  In the three best cases
(SN~2003gd, SN~2005cs, and SN~2008bk), for which multi-filter pre-SN images
exist that enable the SED of the progenitor star to be characterized, the stars
are all found to have been RSG at the lower end of the RSG mass distribution
(i.e., $\lesssim 10\ M_\odot$).  It is noteworthy that the progenitor
characteristics of two of these objects, SN~2005cs \citep{Eldridge07} and
SN~2008bk \citep{Mattila08}, are sufficiently well constrained to be deemed
inconsistent with those expected for massive AGB stars, which are cooler and
significantly more luminous than RSG (see caption to Figure~1).  AGB stars have
been proposed as the direct progenitors of ``electron-capture'' SNe, which
might result from stars at the lower end of the mass sequence ($8\ M_\odot
\rightarrow 10\ M_\odot$?)  triggering collapse of their ONe cores through
electron capture by magnesium-24 and/or neon-20 \citep{Woosley02}.

In the remainder of the progenitor studies of SNe II-P --- with the exception
of two very recent investigations which are, at present, inconclusive (more on
these below) --- the observed or constrained properties of the progenitors are
also consistent with having been RSG at the time of explosion.  This conforms
with expectations (\S~2.2).  However, as shown by \citet{Smartt09a}, a close
look at the data reveals the interesting result that all but one of the $20$
SNe~II-P progenitors have initial masses constrained to be
$\mathrel{\hbox{\rlap{\hbox{\lower4pt\hbox{$\sim$}}}\hbox{$<$}}} 18\ M_\odot$.
In fact, the best fit to the data (assuming a Salpeter IMF of slope $\alpha =
-2.35$, although the result is quite robust to changes in the IMF) yields a
lower mass for SN~II-P progenitors of $M_{\rm min} = 8.5^{+1}_{-1.5}\ M_\odot$
and a maximum mass of $M_{\rm max} = 16.5\pm 1.5\ M_\odot$ \citep{Smartt09a}.
What is more, at this point no progenitor star for {\it any} CC~SN has been
found to have properties consistent with RSG of initial mass $\gtrsim
20\ M_\odot$ (\S~3.3 --- 3.6).  Given that such stars are clearly present in
the Milky Way and Local Group galaxies (see Figure~1), and would have been
easily detected in pre-SN images had they been CC~SN progenitors, the question
arises: What is the fate of the most massive RSG?  A few possibilities to
consider:

\begin{itemize}
\item {\it They do not explode as RSGs}.  Perhaps many (most?  all?)  explode
  during blueward excursions on the HRD.  There is some tentative empirical
  evidence to support this contention.  Two very recent CC~SNe are claimed to
  have had possible YSG progenitors arising from stars with initial with masses
  $\gtrsim 15\ M_\odot$: SN~2008cn \citep{Eliasrosa09} and SN~2009kr
  \citep{Fraser10,Eliasrosa10}.  Definitive identification of both progenitors
  is complicated, however, by potential source confusion since the host
  galaxies are quite distant ($> 30$ Mpc) and the spatial resolution of the
  pre-SN images does not permit distinguishing among single stars, binary
  systems, or compact clusters.  Final progenitor characterization therefore
  awaits late-time imaging.  It is worth noting that neither event appears to
  have been a ``normal'' Type II-P (e.g., in the ilk of SN~1999em;
  \citealt{Leonard5}).  SN~2008cn exhibits spectral peculiarities as well as
  scant published photometric coverage to secure definitive classification as
  an SN~II-P \citep{Eliasrosa09}, whereas SN~2009kr has a photometric evolution
  similar to a Type II-L \citep{Eliasrosa10} or, perhaps, a ``peculiar II-P''
  \citep{Fraser10}.

\item {\it They explode as SNe II-L/IIn/IIb}.  Investigations into this
  possibility are, at present, starved for data.  As we shall see in \S 3.3 ---
  3.5, direct progenitor studies exist for only four SNe II-L/IIn/IIb (not
  including the potential YSG progenitor of SN~2009kr, the possible II-L
  discussed earlier), and in no case is a higher mass RSG implicated.  While
  more examples are clearly needed, at this point there is no evidence that
  high-mass RSG are the direct progenitors of SNe~II-L/IIn/IIb.

\item {\it They do not explode}.  Quiescent collapse to a BH for massive RSG
  remains an intriguing possibility.  As pointed out by \citet{Kochanek08}, at
  this point the optical signatures of direct BH formation are virtually
  unconstrained by either theory or observation.  Could this ``RSG problem''
  (as dubbed by \citealt{Smartt09a}) be the first indication of a mass cutoff
  between stars that successfully eject their envelopes after core collapse
  (i.e., those below $\sim 16.5\ M_\odot$) and stars that do not (i.e., those
  above $\sim 16.5\ M_\odot$)?  We return to this possibility in \S~4.

\end{itemize}

\subsection{Type II-Linear}

Because of their rarity, it is perhaps not surprising that other than the
possible YSG progenitor of the possible Type~II-L SN~2009kr discussed earlier
(\S 3.2), only one additional object, SN~1980K, has had a pre-SN image examined
for a possible progenitor star.  In this case, the analysis rules out massive
RSG greater than about $18\ M_\odot$ \citep{Thompson82}.  Analysis of the
stellar population of the Type II-L SN~1979C by \citet{Vandyk99a} determines a
mass range of $15 - 21\ M_\odot$ for its progenitor.  From these studies, firm
conclusions about the progenitors of SNe~II-L can not be made, although early
indications are that at least some do not arise from extremely massive stars.

\subsection{Type IIn}
There is only one study of a Type~IIn progenitor, and it involves the
interesting case of SN~2005gl.  Initial work by \citet{Galyam07} demonstrated
the spatial coincidence between this SN~IIn and a remarkably bright source with
an estimated luminosity of over $10^6\ L_\odot$, suggesting an LBV progenitor
from luminosity considerations alone (\S~2.1); spectral evidence from SN~2005gl
itself is also consistent with this idea.  Strong claims of an association were
tempered, however, by the distance ($\gtrsim 60$ Mpc) of SN~2005gl's host
galaxy, which translates the $\sim 0.1\arcsec$ resolution of the pre-SN {\it
  HST} image to $\sim 30$ pc, raising suspicion that the detected object could
have been an unresolved stellar cluster or association of several massive
stars, with only part of the light coming from the actual progenitor of
SN~2005gl.  To settle the case, additional {\it HST} observations were obtained
two years later, which clearly demonstrated that the luminous source in the
pre-SN image had disappeared \citep{Galyam09}.  Such a direct association
between an LBV and an SN counters conventional theory (\S 2.1).  Further, an
additional suggestive (but, not conclusive) piece of evidence for such a
connection comes from the peculiar Type Ib SN~2006jc, for which pre-SN images
captured an LBV-like outburst two years prior to the final explosion
\citep{Pastorello07a}.  Unlike SN~2005gl, however, SN~2006jc had clearly lost
its entire hydrogen envelope prior to exploding, and so may have had a
progenitor transitioning from an LBV to a WR.  In both cases, though, the
evidence points towards a more highly evolved core than traditional models
suggest should exist, and may necessitate a rethinking of stellar evolution
theory for Nature's most massive stars \citep[e.g., ][]{Smith06}.

\subsection{Type IIb}
 Pre-SN images exist for two SNe~IIb.  First, SN~1993J in M81, where extensive
 analyses of pre-SN and post-SN images (and spectra) lead to the conclusion
 that a $13 - 20\ M_\odot$ star exploded in a binary system (the only binary
 conclusively implicated thus far by progenitor studies), with a slightly less
 massive secondary surviving the explosion \citep{Maund04,Maund09}.  Second,
 for SN~2008ax, \citet{Crockett08} find a flat SED at optical wavelengths for
 the progenitor object that favors an early-type WR (WN class) progenitor with
 a fairly large ($25 - 30\ M_\odot$) initial mass.  Although late-time images
 are needed to better constrain other possibilities, for now SN~2008ax remains
 the only tentative detection of a WR star progenitor for any CC~SN.

\subsection{Type Ib/c}

There are ten SNe~Ib/c with progenitor studies, but at this point no progenitor
detections.  This is somewhat surprising, since it is commonly thought that at
least some of the progenitors of SNe~Ib/c should be luminous, single WR stars.
While none of the non-detections are sensitive enough to definitively rule out
a WR progenitor, \citet{Smartt09a} demonstrates that it is quite unlikely at
this point that all SNe~Ib/c come from them; lower mass stars in interacting
systems are almost certainly contributing.

\section{Conclusions and Discussion}

Tantalizing clues but few definitive conclusions characterize the infant field
of SN progenitor studies.  A zeroth-order result confirms theory: Only massive
stars appear to explode as CC~SNe.  As of now, there is no evidence that stars
born with significantly less than $ 8\ M_\odot$ undergo core-collapse and
explode.  Another rather firm claim can be made that the direct progenitors of
SNe~II-P are RSGs most likely confined to the lower-mass end of the RSG
population.  Beyond these statements lie outstanding questions for which
present data only hint at resolution.  We conclude by considering one: Do
{\it all} massive stars actually explode at the ends of their lives?

On this point theory provides little guidance, since robust CC explosions
continue to elude modelers.  It may well be that a variety of mechanisms is
required to produce viable explosions across the range of progenitor masses
\citep[e.g., ][]{Kitaura06,Murphy08,Dessart08a}.  In this regard, we take note
of the present tension between evidence that at least some extremely massive
($\gtrsim 25\ M_\odot$) stars do explode (e.g., the progenitor studies of
SN~2005gl, SN~2008ax, and SN~2006jc), with the possibility that some lower mass
stars (i.e., those between $\sim 18\ M_\odot$ and $\sim 25\ M_\odot$ --- the
RSG problem) may not.  Could it be that some version of the standard neutrino
mechanism \citep{Colgate66} is viable for exploding lower-mass stars whereas a
magneto-rotational, jet-induced explosion \citep{Wheeler02} is at work for
those with very high masses, relegating stars that are ``in between'' to quiet
collapse?  

With virtually no effort having been expended on directly searching nearby
galaxies for disappearing massive stars, observation is at present almost as
mute as theory on the issue of quiescent stellar collapse.  Indeed, with a few
notable exceptions\footnote{A limited but otherwise ideal data set for such a
  study is being produced by the author as part of an Archival Legacy {\it HST}
  study (HST-AR-10673; P.I. Leonard) of the repeated observations of nearby
  galaxies observed by {\it HST} for Cepheid studies.}, the requisite data for
an extensive search for ``disappearing stars'' do not yet exist, although
\citet{Kochanek08} makes a compelling case for the need for --- and feasibility
of --- such a survey.  An indirect empirical argument against a significant
number of massive stars failing to explode is made by \citet{Maoz10}, who find
from a distance-limited sample that $0.010 \pm 0.002$ CC SNe are produced per
unit stellar mass formed, a result deemed consistent with expectations provided
{\it all} stars with initial masses greater than $8\ M_\odot$ explode.  As with
statistical arguments made on CC~SNe progenitors in general, though, there is
no substitute for actually looking.

\acknowledgments Support for archival {\it HST} studies is provided by grant
HST-AR-10673 (P.I.: Leonard), administered by NASA through a grant from the
Space Telescope Science Institute, which is operated by the Association of
Universities for Research in Astronomy, Incorporated, under NASA contract
NAS5-26555.  I thank the scientific organizing committee of the HEDLA 2010
conference for inviting this review, and for providing financial assistance to
attend the conference.


%


%

\end{document}